\documentclass[a4paper]{article}
\usepackage{graphicx}
\pdfoutput=1
\newcommand{\be}{\begin{equation}}
\newcommand{\ee}{\end{equation}}
\newcommand{\bd}{\begin{displaymath}}
\newcommand{\ed}{\end{displaymath}}
\newcommand{\bea}{\begin{eqnarray}}
\newcommand{\eea}{\end{eqnarray}}
\newcommand{\bi}{\begin{description}}
\newcommand{\ei}{\end{description}}
\newcommand{\bq}{\begin{quote}}
\newcommand{\eq}{\end{quote}}
\def\i{\item}
\def\fo{\footnote}

\begin{document}
\bibliographystyle{unsrt}%plain
\twocolumn[
\author{Alexander~Unzicker and Julius~Fischer\\
        Pestalozzi-Gymnasium  M\"unchen\\[0.6ex]\\
{\small{\bf e-mail:}  alexander.unzicker@lrz.uni-muenchen.de}}

\author{
Alexander~Unzicker and Julius~Fischer\\
         {\small Pestalozzi-Gymnasium M\"unchen, Germany}\\
           {\small alexander.unzicker@lrz.uni-muenchen.de}}
\title{A Quick Verification of the 2-D Galaxy Distribution  with  SDSS Data}
\maketitle

\begin{abstract}

We present source code for the computer algebra system Mathematica
that analyzes the distribution of nearby Galaxies using SDSS data.
Download instructions are given, thus within 10 minutes, the 
reader can verify that galaxies are distributed in an essentially
non-homogeneous manner and  cluster on 2-dimensional structures.
The short code uses a simple method inspired by Minkowski functionals:
the distances to the next neighbors are calculated and compared to
random distributions in three and two dimensions. The observed 
distance distribution corresponds clearly to the latter case.
The paper may also be helpful for nonexpert scientists to get started with
SDSS data analysis.
\end{abstract}

\vspace{1.0cm}]

\section{Introduction}

Still 50 years after Hubble's discovery of the expanding universe 
the distribution of galaxies was assumed to homogeneous - a seemingly obvious
consequence of the cosmological principle. Pioneering investigations 
\cite{Gel:86, Gel:89} however showed that distribution of 
galaxies is all but homogeneous.
Rather there seems to be a hierarchy of galaxy
groups, clusters and superclusters that concentrate on
twodimensional structures, while there are
large voids in between (`sponge structure', \cite{Got:86}). 
There is an ongoing discussion whether the universe becomes homogeneous
for scales larger than 100 $Mpc$, % yyy check  \cite{}
or if it has the properties of a fractal with $D=2$ even
on larger scales \cite{Syl:08}.

\begin{figure}[h]
\includegraphics[width=8cm]{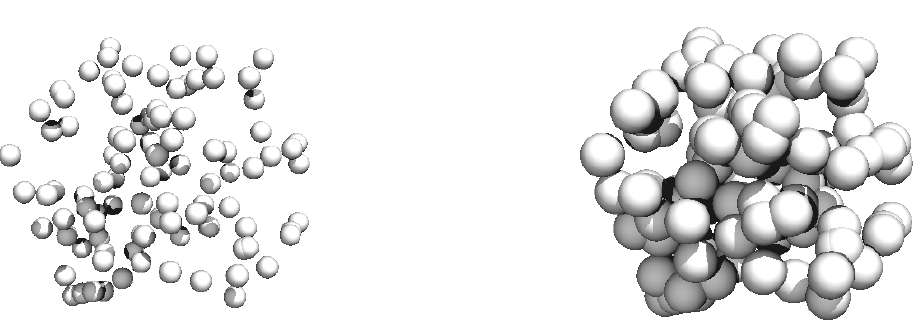}
\caption{Example of Minkowski functionals: at a critical radius, the
manifold becomes connected. Picture taken from \cite{Ost:07}.}
\label{minkowski}
\end{figure}

Here we do not present any new
results that help to decide that question and our approach cannot compete
with the detailedness of the expert's analysis (\cite{Pee:01,Syl:08, Got:08}
and references herein).
From a point of view of general scientific
methodology, we find it however desirable that important 
results of
fundamental physics that require extensive
numerical treatment can be repeated by a 
broad public of non-expert scientists\fo{See, e.g. \cite{Unz:07a} for a similar approach.}.
In particular, the unique quality of the 
free accessible SDSS data supports such an approach
we would like to ease further. Two-point statistics are frequently used
to extract information on the dimensionality. Here we use just next neighbor
statistics. Imagine spheres with growing radius $r$ around each galaxy. When
$r$ reaches a critical radius $r_c$, the spheres will overlap to a 
connected manifold of the size of the whole sample (see fig.~\ref{minkowski}). 

Obviously, this 
transition must occur much earlier (at a small $r_c$) when the distribution
is not homogeneous in three dimensions. We do not fully implement this
method of Minkowski functionals \cite{Mec:94}, but next neighbor distances
obviously do yield significant information.

The code of about 100 lines given below reproduces the results
given in section~\ref{results}. It can easily be run with different data sets
and future data releases of SDSS. We plan to add some refinements 
for a second version, but also the reader should be able to do slight
modifications or extensions of the code. A quick description for
getting started is found in section~\ref{sbs}. Though we cannot
give a detailed description of the program, some clarifying
comments are included in the quite self-explaining code (see
~\ref{sbs2}).

\section{Methods}

\subsection{General method and limitations}
As a first approximative approach, we did not take into consideration
galaxy size and  morphology, which may well influence a more 
refined analysis. Spectra were just used to determine 
the distance by the redshift, $H_0$ was assumed as $72 \ km s^{-1} Mpc^{-1}$
\cite{Rie:05,jac:07}.
Though peculiar velocities cause errors in the radial distance, no correction
was tried so far to take into account that effect. To avoid faint galaxies to
drop out of the sample, we considered redshifts $ z< 0.03$, the point to which
the SDSS data show a roughly constant density. There is a clearly visible decay of
the number of galaxies per volume\footnote{We do not address here the 
question if such a density can reasonably defined for a fractal.} for $D> 130 \ Mpc$ or 
$z > 0.03$ (see fig.~\ref{distrib})\footnote{This picture cannot be generated
by the code given below.}.

\begin{figure}[h]
\includegraphics[width=8.5cm]{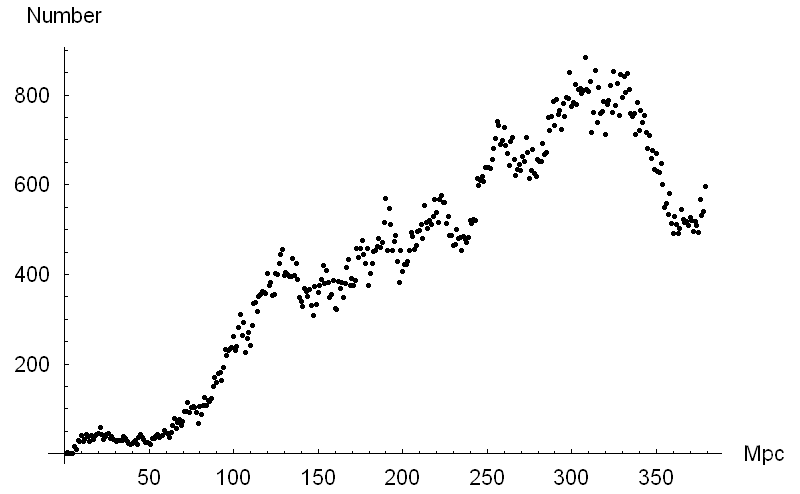}
\caption{Galaxy number between $D$ and $D+dD$ as a function of distance $D$. Constant 
density should lead to a parabolic increase of the number of galaxies 
with distance. For $D> 130 \ Mpc$ or $z >0.03$, obviously a considerable
percentage of galaxies are too faint to be detected. }
\label{distrib}
\end{figure}

\subsection{Data acquisition}

\begin{figure}[h]
\includegraphics[width=8.5cm]{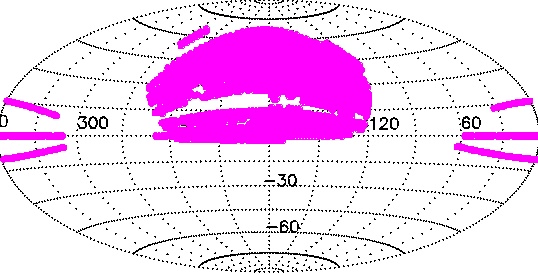}
\caption{Coverage of spectral data from SDSS DR6 site}
\label{DR6cover}
\end{figure}

\begin{figure}[h]
\includegraphics[width=7cm]{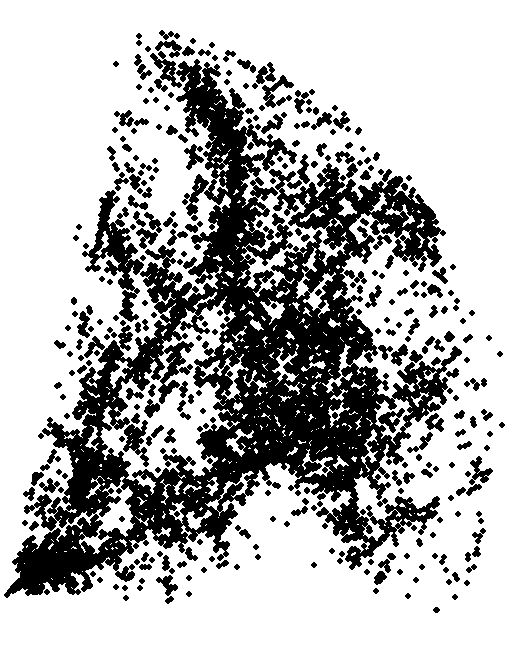}
\caption{3D-plot of the position of the 9280 galaxies of sample 1.}
\label{galpl}
\end{figure}

The sixth data release (DR6) of the SDSS data is located in
$http://www.sdss.org/dr6$ (see fig.~\ref{DR6cover}). Though very simple data sets can be accessed
by search masks, we strongly recommend the use of SQL data search language
to which the SDSS site provides very good tutorials. 
The part of the sky taken into consideration was determined by the SDSS coverage. We
chose $140^{\circ} < DEC 240^{\circ} $ and $30^{\circ} < RA 60^{\circ}$ and a second 
sample $140^{\circ} < DEC 240^{\circ} $ and $-2^{\circ} < RA 11^{\circ}$.  See 
fig.~\ref{galpl} for a 3D-plot of the galaxies of sample 1. 
The confidence level was set to $0.35$. The respective SQL commands for downloading
the data are listed in the appendix~\ref{sbs}.

\subsection{Data manipulation and modelling}

We used the computer algebra system $Mathematica$ to convert the raw
data to Euclidean coordinates. Thus the distances to the next neighbor 
could be calculated easily. Then, these minimal or next neighbor distances 
for each galaxy contain the desired structure information.
For a large number of $i$ points, an effective minimal distance algorithm is needed,
since computing {\em all\/} distances would lead to an increase of computational 
time $t \sim i^2$. We chose a very simple method that leads to
$t \sim i$. The entire volume was divided by a rectangular lattice in $n^3$ 
boxes of equal size\footnote{for the real distribution, an equal angular size was used.},
e.g. for $n=10$ into 1000 boxes. In a first step, each galaxy was assigned to its box.
To determine the minimal distance of an individual galaxy, the computation of {\em all\/}
distances within one box and the 26 neighboring boxes was sufficient\footnote{Of course,
in 2D there are 8 neighboring boxes.}. % yyy As a little Randeffect, 
Likewise, the random distributions were analyzed, whereby in the 2D-case a $n^2$-lattice 
with larger $n$ was chosen.\footnote{$n$ does however influence the computational time only.}
The 3D- simulation consisted of a cube with the same volume as the real sample shown
in fig.~(\ref{galpl}). The galaxy density was equal to the real one. For the
2D-simulation, the size of the surface of a sphere was chosen, while the spherical
volume was equal to the real one. Due to computational simplicity, the form of the
surface was a square.

\section{Results - a preliminary analysis}
\label{results}

\begin{figure}[h]
\includegraphics[width=8.5cm]{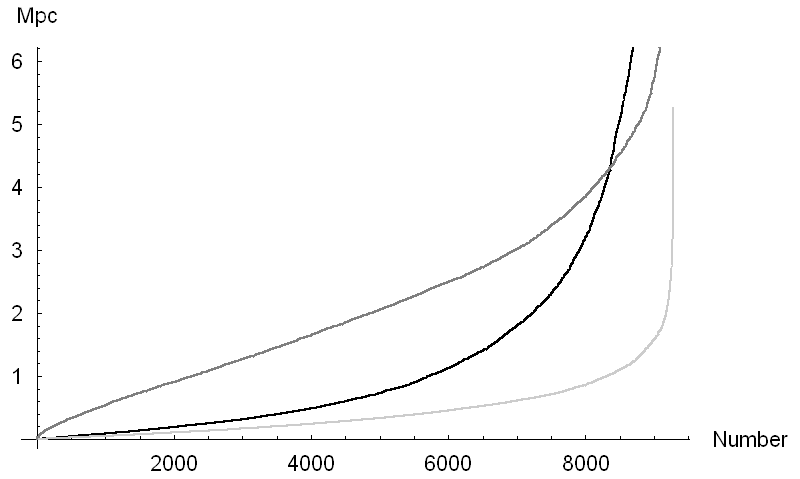}
\caption{Sorted next-neighbor-distances, for the real distribution (black) 
and the simulations in 3-D (dark gray)
and 2D (light gray). Sample from the region $ 60^{\circ} > DEC > 30^{\circ}$ 
and $ 240^{\circ} > RA > 140^{\circ}$.}
\label{sorted}
\end{figure}

\begin{figure}[h]
\includegraphics[width=8.5cm]{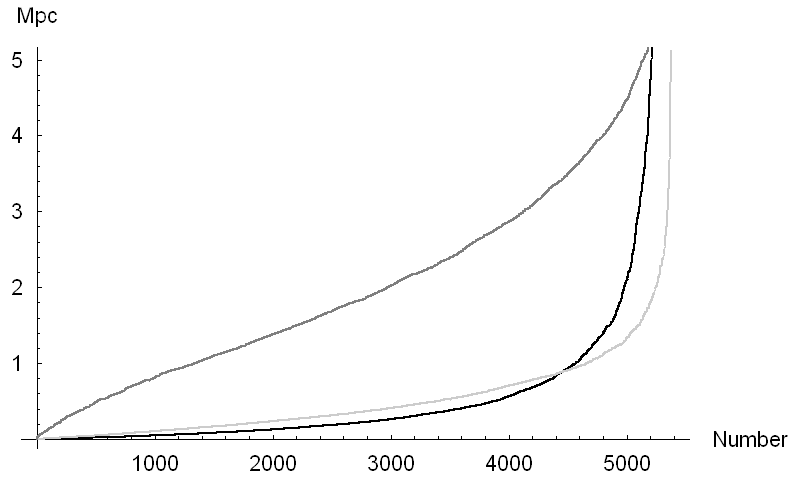}
\caption{as fig.~\ref{sorted}, but for a smaller sample $ 11^{\circ} > DEC > -2^{\circ}$ 
and $ 240^{\circ} > RA > 140^{\circ}$.}
\label{sorted2}
\end{figure}

The real sorted distances (fig.~\ref{sorted} and \ref{sorted2}) exclude clearly a 
homogeneous 3D-distribution which would lead 
to greatly different next-neighbor-distances. They also coincide well with the 2D random
distribution on a surface, as far as small distances are concerned. The approach to the
3D-simulation for larger distances could indicate a homogeneity of the universe on larger 
scales. We do not know how  peculiar velocities  influence, the
effect should however be limited to $D < 5 \ Mpc$ (\cite{Syl:08}, p.~6)
Next neighbors % leo
which occasionally are to faint to be detected could as well spoil the
analysis at larger distances. Especially the presence of large voids in the real
distribution may explain why there the largest next-neighbor-distances exceed those
of the random distribution.
 Therefore, as mentioned in the introduction, an interpretation
of the present data in favor of a large-scale homogenity instead of a fractal dimension
$D=2$ would be premature.
We have no explanation so far for the difference visible in the two samples 
fig.~(\ref{sorted}) and (\ref{sorted2}). 

\section{Conclusions}

The distribution of galaxies in the universe is still a riddle and theoretically 
not fully understood \cite{Pee:01}. The discovery of the two-dimensional structure 
of the galaxy distribution reminds us from something deep and mysterious,
such as from Dirac's observation of a twodimensional density in the 
universe.\footnote{From Dirac's large number hypothesis follows that the surface
of all protons is of the same order as the surface of the horizon. Today this is
usually considered as pure coincidence.}
Particulary $\Lambda$CDM simulations have problems to account for the observed
structure. Observational progress in this field is therefore very much 
triggered by new high-quality data like SDSS. The ongoing discussion can thus benefit
from a broad accessibility of those data and a transparent processing which is not
limited to a few groups. If one day the data allow a definite answer to decide whether
the large scale structure is homogeneous of even fractal, this must become evident
also for the non-expert scientist. We hope this is a little step towards
repeatability and transparency for the efforts to answer that important question.

\paragraph{Acknowledgement.} Though we are grateful for any comments, please 
understand that we cannot guarantee functionality or give further 
support for getting this program to run on your computer.

\onecolumn

\section{Appendix:  data preparation and source code}

\subsection{Step-by step procedure in 10 minutes}
\label{sbs}
\begin{enumerate}
\i Create your directory `sdss' and copy all the following files in there.
\i Browse to $http://www.alexander-unzicker.de/sdss1.txt$ and copy the file. Alternatively,
copy and paste from the arXiv source and save as $sdss1.txt$. 
\i Browse to $http://cas.sdss.org/dr6/en/tools/search/sql.asp$
\i Type the following SQL commands in the blank field (you may paste and
copy it also from the end of the $sdss1.txt$ file):
\end{enumerate}
\begin{verbatim}
select ra,dec,z
from specObj
where ra BETWEEN 140 and 240 AND
dec BETWEEN 30 and 60 AND 
specClass = 2 AND
z BETWEEN 0.001 AND 0.03
AND zConf > 0.35
\end{verbatim}
\begin{enumerate}
\setcounter{enumi}{4}
\i Chose file format CSV.
\i Press submit and save the data file as $sdss03.csv$.
\i Proceed likewise for dec between -2 and 11 and save as  $sdss03a.csv$.
\i Open a Mathematica *.nb file and run the following commands (apart from the
SetDirectory comand where you have to put in {\em your\/} path, you may paste and
copy it also from the end of the $sdss1.txt$ file)
\end{enumerate}
\begin{verbatim}
SetDirectory["yoursdsspath"];
<<"sdss1.txt";
readData["sdss03.csv"];  (* later sdss03a.csv *) 
GalPlot[{100, 500, 700}];
fastDistances[rpt, xyz, {20, 20, 20}, {rarg, decrg, rrg}];
randomDistances3d[vol^(1/3), Galnumber, {20, 20, 20}];
randomDistances2d[(3 vol/4/Pi)^(1/3) Sqrt[4 Pi], Galnumber, {90, 90}];
compareDistances[0.004]; (** creates one plot from 3  ***)
\end{verbatim}

\subsection{Source code}
\label{sbs2}
\begin{verbatim}
<<Statistics`DataManipulation`;
(************** plot options *************)
sty2={{GrayLevel[0.3],Thickness[0.01]},{GrayLevel[0.5],
Thickness[0.01]},{GrayLevel[0.8],Thickness[0.01]}}; 
(********************* constants ************************)
cc=299792458; H0=1/(4.4081 10^(17)); Mpc=3.08567758128*10^(22);(*corresponds to 70 km/s/Mpc*)
offset = 0.00000001;(**** to avoid division by zero ****)
(************* functions needed ************************)
tor[z_]:=(2cc z+cc z^2)/(H0 Mpc(2+2z+z^2));(* transformation from redshift to Mpc distance *) 
toxyz[{r_,th_,ph_}]:={r Sin[th]Cos[ph],r Sin[th]Sin[ph],r Cos[th]}; (*to cartesian coord.*)
dist2[k1_,k2_]:=Apply[Plus, (k1-k2)^2];(*square of distance of two points k1, k2 in 3 dim.*)
(*************** procedures**********************************)
(********* reads SDSS data in *.csv format  *****************)
readData[infile_]:=Block[{qwe,wer,wer1,ra,dec,rr, zz}, (* local variables *)
If[FileInformation[infile]=={},Print["file not found."]; Goto[endlabel]]; (* check infile *)
qwe = Drop[Import[infile, "CSV"], 1]; wer1 = Transpose[qwe];
wer = ReplacePart[wer1, offset, Position[wer1, 0]];(* replace undesired zeros *)
ra = 2 Pi wer[[1]]/360; dec=2 Pi wer[[2]]/360; zz=wer[[3]]; (* angle in radians *)
(*** automatic boundary determination from data , ranges of coordinates ****)
{ramin, decmin}={Min[ra]-offset,Min[dec]-offset}; 
{ramax, decmax}={Max[ra]+offset,Max[dec]+offset}; 
zmin=Min[zz]-offset; zmax=Max[zz]+offset; rarg=ramax-ramin; decrg=decmax-decmin;
zrg=zmax-zmin; meandec=(decmax+decmin)/2; 
rrmax = tor[zmax]; rrmin = tor[zmin]; rrg = rrmax - rrmin;
rr = Map[tor, zz]; (*** transformation redshift - radius**)
rpt=Transpose[{rr, dec, ra}]; (*spherical coordinates*)
xyz= Map[toxyz, rpt];(* transform to cartesian coordinates *)
vol=(rrmax^3-rrmin^3) 4/3 Pi rarg/(2Pi) decrg/(2Pi) Abs[Cos[meandec]];(* mind latitude *)
Galnumber=Length[rpt]; GalDichte=Galnumber/vol;
Print["Number of galaxies: ", Galnumber]; Print["Volume in Mpc^3: ", vol];
Label[endlabel]];
(************** calculates real distances *************************)
realDistances[rpt_,xyz_,unt_,rgs_]:=Block[{}, tu1=TimeUsed[]; 
{rarg, decrg, rrg}=rgs; (* ranges*)
box = mindist = Table[{}, {unt[[1]]}, {unt[[2]]}, {unt[[3]]}];
(* empty variable for the boxes and minimal distances*)
(* assign each galaxy to its box *)
(* though boxes are defined by polar coordinates, they contain cartesian ones *)
For[i = 1, i <= Galnumber, i++, AppendTo[box[[Ceiling[(rpt[[i, 1]] - rrmin)/rrg *unt[[1]]],
   Ceiling[(rpt[[i, 2]]-decmin)/decrg*unt[[2]]],
   Ceiling[(rpt[[i,3]]-ramin)/rarg *unt[[3]]]]],xyz[[i]]]];
tu2=TimeUsed[];Print["assign to boxes... ",tu2-tu1, "s"];
(**** distance in box i,j,k to all other galaxies in the box and in neighboring boxes ******)
For[i = 1, i <= unt[[1]], i++, 
  For[j = 1, j <= unt[[2]], j++,
    For[k = 1, k <= unt[[3]], k++, 
        For[m = 1, m <= Length[box[[i, j, k]]], m++, distances={};
          (*now go through neighboring boxes ii,jj, kk *)
  For[ii = i-1, ii <=i+1, ii++,  
    For[jj = j-1, jj <= j+1, jj++,
      For[kk = k-1, kk <= k+1, kk++, 
        If[(ii==0 || jj==0 || kk==0 || ii==unt[[1]]+1 || jj==unt[[2]]+1 || kk==unt[[3]]+1), 
            Continue, 
   If[box[[ii,jj,kk]]!={},AppendTo[distances,Table[dist2[box[[i,j,k,m]],box[[ii,jj,kk,mm]]],
      {mm,Length[box[[ii, jj, kk]]]}]]]]; ]]];  (* avoid pathologic cases like empty boxes *)
         If[(distances!={} && Flatten[distances]!={0.}),
            AppendTo[mindist[[i,j,k]],Sort[Flatten[distances]][[2]]]]; 
]]]]; 
tu1=TimeUsed[];Print["measuring...",tu1-tu2, "s"];
distToNext = Flatten[mindist]; rowOfDist=Sort[distToNext]; 
ListPlot[rowOfDist,PlotJoined ->True, AxesLabel -> {"Number", "Mpc"}]];
(******* calculates distances of random distibutions in 3D  *****************)
randomDistances3d[edge_,nn_,unt_]:=
Block[{box,mindist,distances,distToNext, i,j,k,m}, tu1=TimeUsed[];
xyz=Table[edge{Random[],Random[],Random[]},{i,nn}];
box = mindist = Table[{}, {unt[[1]]}, {unt[[2]]}, {unt[[3]]}];
(* empty variable for the boxes and minimal distances*)
(* assign each galaxy to its box *)
For[i = 1, i <= nn, i++, AppendTo[box[[Ceiling[(xyz[[i, 1]])/edge*unt[[1]]],
 Ceiling[(xyz[[i, 2]])/edge*unt[[2]]], Ceiling[(xyz[[i,3]])/edge*unt[[3]]]]], xyz[[i]]]];
tu2=TimeUsed[];Print["assign to boxes...  ",tu2-tu1, "s"];
For[i = 1, i <= unt[[1]], i++, 
   For[j = 1, j <= unt[[2]], j++,
     For[k = 1, k <= unt[[3]], k++, 
         For[m = 1, m <= Length[box[[i, j, k]]], m++, distances={};
         (*now go through neighboring boxes ii,jj, kk *)
 For[ii = i-1, ii <=i+1, ii++,  
    For[jj = j-1, jj <= j+1, jj++,
      For[kk = k-1, kk <= k+1, kk++, 
        If[(ii==0 || jj==0 || kk==0 || ii==unt[[1]]+1 || jj==unt[[2]]+1 || kk==unt[[3]]+1),
           Continue, 
             If[box[[ii,jj,kk]]!={},AppendTo[distances, Table[dist2[box[[i,j,k,m]],
                 box[[ii,jj,kk,mm]]],{mm,Length[box[[ii, jj, kk]]]}]]]];
            ]]];
            If[(distances!={}&& Flatten[distances]!={0.}),
               AppendTo[mindist[[i,j,k]],Sort[Flatten[distances]][[2]]]];
]]]];
tu1=TimeUsed[];Print["measuring...",tu1-tu2, "s"];
distToNext = Flatten[mindist];
rowOfDist3=Sort[distToNext];
ListPlot[rowOfDist3,PlotJoined ->True, AxesLabel -> {"Number", "Mpc"}]];
(******* calculates distances of random distibutions in 2D  *****************)
randomDistances2d[edge_,nn_,unt_]:=
Block[{box,mindist,distances,distToNext, i,j,m}, tu1=TimeUsed[];
xy=Table[edge{Random[],Random[]},{i,nn}];
box = mindist = Table[{}, {unt[[1]]}, {unt[[2]]}];
(* empty variable for the boxes and minimal distances*)
(* assign each galaxy to its box *)
For[i = 1, i <= nn, i++, AppendTo[box[[Ceiling[(xy[[i, 1]])/edge*unt[[1]]],
 Ceiling[(xy[[i, 2]])/edge*unt[[2]]]]], xy[[i]]]];
tu2=TimeUsed[];Print["assign to boxes...  ",tu2-tu1, "s"];
For[i = 1, i <= unt[[1]], i++,
  For[j = 1, j <= unt[[2]], j++,
     For[m = 1, m <= Length[box[[i, j]]], m++, distances={};
          (*now go through neighboring boxes ii,jj *)
 For[ii = i-1, ii <=i+1, ii++,
   For[jj = j-1, jj <=j+1, jj++,
      If[(ii==0 || jj==0 || ii==unt[[1]]+1 || jj==unt[[2]]+1), Continue, 
      If[box[[ii,jj]]!={},AppendTo[distances, Table[dist2[box[[i,j,m]],
           box[[ii,jj,mm]]],{mm,Length[box[[ii, jj]]]}]]]];
       ]];
        If[(distances!={}&& Flatten[distances]!={0.}),
            AppendTo[mindist[[i,j]],Sort[Flatten[distances]][[2]]]];
]]];
tu1=TimeUsed[];Print["measuring...",tu1-tu2, "s"];
distToNext = Flatten[mindist];
rowOfDist2=Sort[distToNext];
ListPlot[rowOfDist2,PlotJoined ->True, AxesLabel -> {"Number", "Mpc"}]];
(***************** plot of galaxy distribution **************************)
GalPlot[vp_(*chose appropriate ViewPoint*)]:=Show[Graphics3D[Map[Point, xyz]],
Boxed->False, ViewPoint -> vp, Prolog->AbsolutePointSize[0.0005]];
compareDistances[th_]:=Block[{},$DefaultFont={"Arial", 8};
lp1 = ListPlot[rowOfDist, PlotStyle -> {Thickness[th],GrayLevel[0]}, 
PlotJoined->True, DisplayFunction -> Identity];
lp2 = ListPlot[rowOfDist2, PlotStyle ->{Thickness[th],GrayLevel[0.8]}, 
PlotJoined->True,  DisplayFunction -> Identity];
lp3 = ListPlot[rowOfDist3, PlotStyle -> {Thickness[th],GrayLevel[0.5]}, 
PlotJoined->True,   DisplayFunction -> Identity];
distrib=Show[lp1,lp2,lp3, DisplayFunction -> $DisplayFunction, 
AspectRatio -> .6, AxesLabel -> {"Number", "Mpc"}]];
\end{verbatim}
\end{document}